\documentclass[%
 reprint,
 amsmath,amssymb,
 aps,
 prb,
 floatfix,
 footnoteinthebib,
 longbibliography
]{revtex4-2}
\usepackage{nicematrix,tikz}
\usepackage{color}
\usepackage{dcolumn}
\usepackage{latexsym}

\usepackage[normalem]{ulem}
\usepackage{hyperref,amssymb}
\usepackage{url}
\usepackage{color,soul}
\usepackage{graphicx}
\usepackage{verbatim}
\usepackage{multirow}
\usepackage{amsmath}
\newcommand{\beq}{\begin{eqnarray}}
\newcommand{\eeq}{\end{eqnarray}}
\usepackage{mathrsfs}

\usepackage[latin1]{inputenc}

\begin{document}

\title{\Large Revisiting the phonon theory of liquid heat capacity:\\ low-frequency shear modes and intramolecular vibrations}
\author{Yu Liu$^{1}$}
\email{liu_yu@cqu.edu.cn}
\author{Matteo Baggioli$^{2,3,4}$}
\email{b.matteo@sjtu.edu.cn}
\address{$^1$Key laboratory of low-grade Energy Utilization Technologies and systems,
Ministry of education, School of energy and power Engineering, Chongqing
University, Chongqing, 400030, China}
\address{$^2$School of Physics and Astronomy, Shanghai Jiao Tong University, Shanghai 200240, China}
\address{$^3$Wilczek Quantum Center, School of Physics and Astronomy, Shanghai Jiao Tong University, Shanghai 200240, China}
\address{$^4$Shanghai Research Center for Quantum Sciences, Shanghai 201315,China}

\begin{abstract}
Modeling the heat capacity of liquids present fundamental difficulties due to the strong intermolecular particle interactions and large diffusive-like displacements. Based on the experimental evidence that the microscopic dynamics of liquids closely resemble those of solids, a phonon theory of liquid thermodynamics has been developed. Despite its success, the phonon theory of liquids relies on the questionable assumption that low-frequency shear excitations are propagating in nature and follow a Debye density of states. Furthermore, the same framework does not capture the contribution of intramolecular vibrations, which play a significant role in molecular liquids. In this work, we revisit the phonon theory of liquid heat capacity, introducing alternative approaches to model low-frequency shear modes. In particular, we consider the recently proposed idea of treating such modes as pure kinetic and we propose a novel approach based on identifying those low-frequency excitations as overdamped liquid-like modes with linear in frequency density of states. Moreover, we complete the theory by incorporating the effects of intramolecular vibrations. By comparing the theoretical predictions from these different approaches with the available data for the heat capacity of several liquids, we present a comprehensive evaluation of the original model and the newly proposed extensions. Despite all approaches perform well at low-temperatures, our results indicate that modeling low-frequency modes as overdamped liquid-like excitations yields the most accurate agreement with the data in the whole temperature range. Conversely, we demonstrate that treating these excitations as purely gas-like leads to significant inaccuracies, particularly at high temperatures.

\end{abstract}

\maketitle

\section*{Introduction}
Heat capacity (or specific heat) $c_v$ is a fundamental thermodynamic property of matter. Among the various phases of matter, the heat capacity of (crystalline) solids and gases, and in particular its temperature dependence, are well-understood and indeed successfully described by phonon (Debye) theory \cite{ashcroft1976solid} and kinetic theory \cite{kauzmann2012kinetic}. Using these formalisms, it is straigthfroward to derive for example that the heat capacity of a monoatomic gas equals $c_v=3 R/2$, with $R$ the ideal gas constant. On the other hand, one can also obtain that the heat capacity of harmonic crystals at high-temperature (\textit{i.e.}, in the classical limit above the Debye temperature) is given by the Dulong-Petit law, $c_v=3 R$.

\begin{figure}
    \centering
\includegraphics[width=\linewidth]{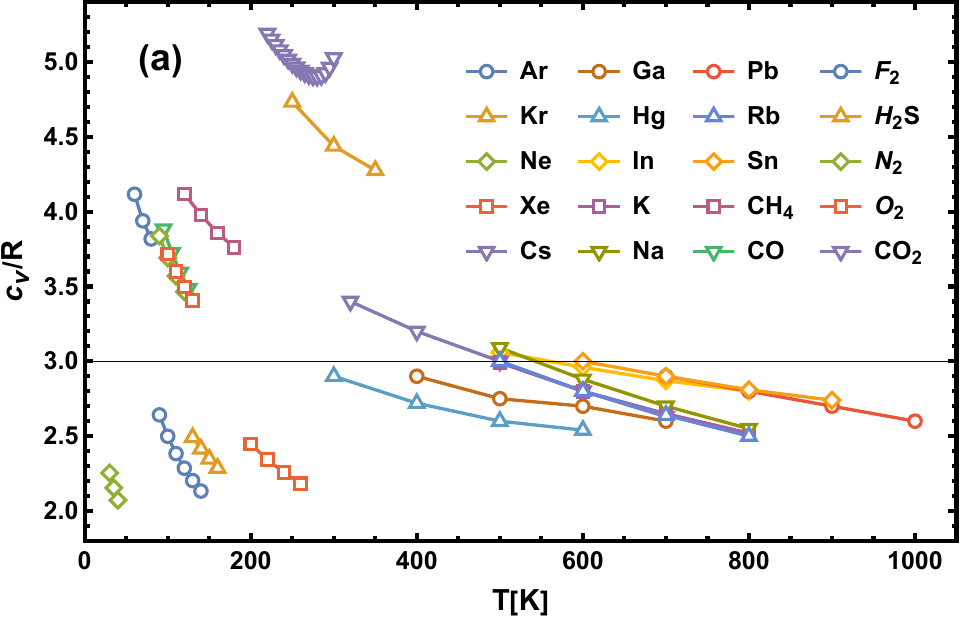}

\vspace{0.2cm}

\includegraphics[width=\linewidth]{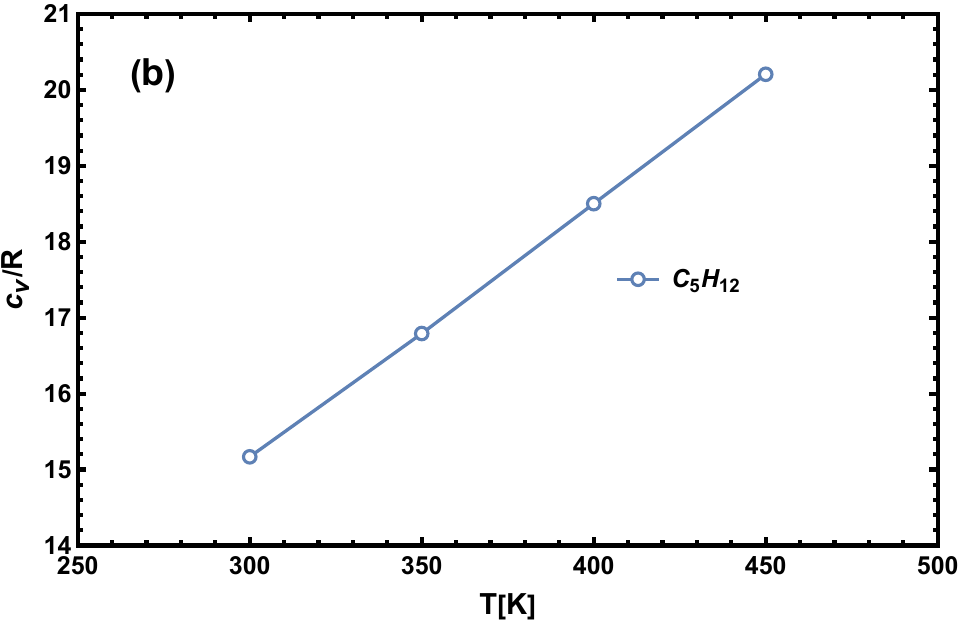}
    \caption{\textbf{(a)} Heat capacity of several liquids as a function of temperature $T$. Data are taken from the NIST REFPROP database \cite{nist}. Panel \textbf{(b)} shows the case of n-pentane (C$_5$H$_{12}$) as a representative liquid for which $c_v$ grows monotonically with $T$.}
    \label{fig1}
\end{figure}

However, when it comes to liquids, these formalisms incur in several difficulties, leaving the theoretical modeling of the liquid heat capacity as an open question (see for example \cite{PhysRevB.78.104201}). Panel (a) in Fig. \ref{fig1} shows the heat capacities of several typical liquids taken from the NIST REFPROP database \cite{nist}. Overall, the larger the molecular weight, the greater the heat capacity. Moreover, the specific heat of molecular liquids is generally higher than that of monatomic liquids. In addition, it can be observed that for liquids with simple structure, such as Ar, CO, CH$_4$, their heat capacities show a decreasing trend with temperature, opposite to the typical behavior in solids. This trend could have been anticipated by extrapolating $c_v$ between the high-temperature solid value ($c_v=3 R$) and the gas-like limit ($c_v=3/2 R$). Nevertheless, there are some fluids, like CO$_2$, for which their heat capacity decreases firstly and then goes up with temperature. Finally, as shown in panel (b) of Fig. \ref{fig1}, more complex liquids, such as n-pentane (C$_5$H$_{12}$), display a heat capacity monotonically increasing with temperature. In summary, liquids with different structures exhibit different dependencies of the heat capacity on temperature, which makes it difficult to propose a general model for the heat capacity of liquids.

Currently, there are different approaches for studying the heat capacity of liquids. One of them approximates liquids with gas-like models focusing on the equation of state (EoS) \cite{sengers2000equations,10.1063/1.555898} or using massive molecular dynamics (MD) simulations \cite{B517931A}. The equation of state is, in fact, the most conventional method to calculate liquid heat capacity. It is usually expressed as an empirical relation between temperature and pressure, where the empirical parameters are determined by fitting the experimental data. Despite the computational accuracy of this method could be satisfactory, this approach leaves the physical picture completely unclear. Furthermore, the extrapolation performance cannot be guaranteed in high temperature/pressure regions where experimental data are unavailable. As regard to molecular dynamics, due to the strong intermolecular interactions in liquids and their significant dependence on the specific molecular structure, it is very complex and difficult to solve this problem strictly based on interactions and correlation functions. In practical applications, simple forms for the particle interactions (such as pair potential) are often applied to reduce computational difficulty, but this often results in inaccurate predictions. For instance, the uncertainty in calculating the heat capacity of C$_3$H$_2$F$_4$ can be as high as $30$\% \cite{doi:10.1021/jp102534z}. 

A second interesting approach to model liquid thermodynamics is the expression of the excess internal energy and specific heat proposed by Rosenfeld and Tarazona \cite{ROSENFELD10101998} using density functional theory and perturbation theory. This method predicts a power-law decay of the excess isochoric heat capacity $c_v^{ex} \sim T^{-2/5}$, which is in good agreement with experimental data, specially for liquids with strong correlations between equilibrium fluctuations of virial and potential energy \cite{10.1063/1.4827865}. Still this path does not make any reference to the degrees of freedom responsible for the heat capacity of liquids, leaving the physical mechanism behind liquid heat capacity undetermined. Moreover, many systems display a rather different power-law decay (see \textit{e.g.} \cite{10.1063/5.0230219}) that is not captured by the model of Rosenfeld and Tarazona.

Another successful direction is based on the so-called two-phase model of liquid thermodynamics \cite{10.1063/1.1624057,C0CP01549K} that idealizes a liquid as a superposition of gas-like and solid-like degrees of freedom. The main difficulty in this model is the determination of the \textit{fluidity parameter} that determines the ratio between the number of these two types of degrees of freedom. We notice that this idea has been recently revisited using instantaneous normal modes \cite{PhysRevE.108.014601}, deriving this fluidity parameter from the fraction of stable and unstable modes (see also \cite{moon2024heat} for a recent book on the nature of `\textit{normal modes}' in liquids). At the same time, unstable instantaneous normal modes have been proposed as the microscopic origin of the temperature decay of the liquid heat capacity as well \cite{PhysRevE.104.014103}.

Finally, another approach, and the primary focus of this work, involves studying liquids from a solid-state perspective by adapting the phonon theory of solids to liquids (see \cite{trachenko2023theory} for a comprehensive historical overview). The core idea of this model lies in the shared characteristics of collective excitations in both liquids and solids. Specifically, the phonon theory of liquid thermodynamics \cite{PhysRevB.78.104201,PhysRevB.84.054106,Bolmatov2012} attributes the temperature-dependent decrease in $c_v$ to the gradual disappearance of collective shear waves in liquids. This behavior is captured by the so-called $k$-gap dispersion relation, $\mathrm{Re}(\omega_s) = v_s \sqrt{k^2 - k_g^2}$, where $\omega_s$ is the frequency of shear waves, $v_s$ is their high-frequency velocity, and $k_g$ is the $k$-gap, defined as $k_g = 1 / (2 v_s \tau)$. Here, $\tau$ is commonly identified with the Maxwell relaxation time $\tau_M$ and decreases quickly with temperature. The $k$-gap dispersion relation has been validated through molecular dynamics simulations \cite{PhysRevLett.118.215502} and experimentally observed in dusty plasma \cite{PhysRevLett.97.115001} and granular fluids \cite{jiang2024experimental}. For additional details, we refer to \cite{Trachenko_2016,BAGGIOLI20201}. In a nutshell, while $\tau$ decreases with temperature, long-wavelength shear waves progressively disappear (or better, become overdamped), drastically reducing the contribution of collective shear modes to the liquid heat capacity.

Importantly, the phonon theory of liquid heat capacity demonstrates strong agreement with experimental and simulation data. Moreover, its particular strength lies in the simplicity and generality, as it requires no free-fitting parameters. We refer to \cite{10.1063/5.0025871} for an extensive comparison between this theory and experimental results (see also \cite{doi:10.1021/acs.jpclett.2c01779} for its potential applications to complex liquids and biological fluids). Nevertheless, we anticipate that the phonon theory of liquid thermodynamics present two fundamental shortcomings that will constitute the main topic of this work.

First, the phonon model for liquid heat capacity has been applied only to simple liquids whose heat capacities decrease with temperature. More precisely, the model cannot capture a situation in which the liquid heat capacity increases with temperature, as for more complex liquids like C$_5$H$_{12}$ (see panel b in Fig. \ref{fig1}). This deficiency comes from the assumption of neglecting intramolecular vibrations, whose effects become important, and even dominant, in certain situations.

Second, and more fundamentally, the phonon model for liquid heat capacity is based on an incorrect physical assumption on the nature of the low-frequency shear modes. In particular, in order to estimate the kinetic energy contribution for shear modes below the so-called Frenkel frequency $\omega_F \equiv 1/\tau$, it is assumed that such excitations are propagating waves following a Debye density of states. As clearly recognized in the recent review by Chen \cite{chen2022perspectives}, this is highly questionable since there are no propagating underdamped shear waves below such frequency scale. More precisely, as also recently emphasized in \cite{brazhkin2024density}, those excitations correspond rather to liquid-like overdamped modes. We notice that this question is tightly connected to whether the low-frequency `density of states' (DOS) in liquids is of Debye form. In fact, that is not the case, as demonstrated by several experimental works \cite{PhysRevLett.63.2381,DAWIDOWSKI2000247,doi:10.1021/acs.jpclett.2c00297,Jin2024} (see also \cite{Ding2025}). A linear in frequency scaling regime of the low-energy liquid DOS is indeed observed and often rationalized with the presence of overdamped relaxational modes \cite{zaccone2021universal}. As revisited below, the same linear behavior can be directly extracted from the gapped dispersion relation of collective shear waves in liquids as well \cite{Trachenko_2023,yu2023unveiling,brazhkin2024density}.

The purpose of this work is to revisit these two shortcomings of the otherwise successful phonon model of liquid thermodynamics. While fixing the absence of intramolecular vibrational modes is rather straightforward, the second and more fundamental question presents a more difficult challenge. There, we will (I) consider the proposal of consider those low-frequency shear modes as gas-like and purely kinetic \cite{chen2022perspectives} and (II) we will propose a way to treat them as liquid-like overdamped modes, more consistent with their real physical nature. We will then compare the original model and these two new extensions to the experimental data and explore in details the problems of each approach. In conclusion, the model that treats low-frequency shear modes as liquid-like will result as the most accurate in predicting the heat capacity of liquids over a large range of temperatures.

\section*{Theory}
In this section, we will revisit the original phonon model for liquid thermodynamics \cite{Bolmatov2012} and introduce several extensions aimed at addressing its limitations and enhancing its predictive accuracy.

As now widely accepted, liquids involve two types of motion: molecular vibrations around the equilibrium
positions (\textit{akin} to phonons in solids) and diffusional motion between adjacent positions, that ultimately leads to their \textit{fluidity}. Hence, the total energy $E$ of a liquid can be decomposed as
\begin{equation}
    E= E_v+E_d,
\end{equation}
where $E_v$ and $E_d$ are respectively the vibrational and diffusional energy. In addition to the diffusional motion, liquids support longitudinal waves with all frequencies and shear waves with
frequencies exceeding the Frenkel frequency $\omega_F\equiv 2\pi/\tau$ \cite{Frenkel1946}. The timescale $\tau$ is usually identified with the Maxwell relaxation time $\tau_M=\eta/G_{\infty}$ with $\eta,G_\infty$ respectively the shear viscosity and the infinite-frequency (or instantaneous) shear modulus \cite{Trachenko_2016,BAGGIOLI20201}. This is the starting point of the phonon theory of liquid thermodynamics. 

One can then further decompose the total energy of a liquid as,
\begin{equation}
    E=K_l+P_l+K_s(\omega>\omega_F)+P_s(\omega>\omega_F)+K_d+P_d, \label{eq2}
\end{equation}
where $K$ and $P$ are kinetic and potential components of the energy. Subscript $l$, $s$ and $d$ represent
longitudinal, shear and diffusion, respectively. It has been argued in \cite{Bolmatov2012} that $P_d$ in Eq. \eqref{eq2} is significantly smaller than the other terms, rendering it negligible. Following that simplifying assumption, and combining the kinetic energy terms, the total energy can be
expressed as below,
\begin{equation}
    E=K+P_l+P_s(\omega>\omega_F),
\end{equation}
where $K$ stands for the total kinetic energy of the system.
By imposing the energy equipartition theorem, we obtain
\begin{align}
    P_l=\frac{E_l}{2},\quad P_s(\omega>\omega_F)=\frac{E_s(\omega>\omega_F)}{2},
\end{align}
and
\begin{equation}
    K=K_l+K_s=\frac{E_l}{2}+\frac{E_s}{2}.
\end{equation}
Since $E_s=E_s(\omega<\omega_F)+E_s(\omega>\omega_F)$, the total energy is then
\begin{equation}\label{toto}
E=E_l+E_s(\omega>\omega_F)+\frac{E_s(\omega<\omega_F)}{2}.
\end{equation}
Up to this point, our treatment coincides exactly with the original phonon theory of liquid thermodynamics \cite{Bolmatov2012}.

To proceed, the Authors of Ref.~\cite{Bolmatov2012} consider the canonical phonon free energy and assume a Debye density of states for all these modes. As also noted by Chen \cite{chen2022perspectives}, this last assumption for the contribution of the shear modes with $\omega<\omega_F$ is problematic, and indeed highly questionable from a physical point of view. In fact, as recently also emphasized by Brazkhin \cite{brazhkin2024density}, shear modes with $\omega<\omega_F$ are non-propagating localized excitations corresponding to overdamped and non-oscillatory motion. Because of their nature, it is not physically justified to treat them as propagating waves or harmonic oscillators, as done in \cite{Bolmatov2012}. Moreover, the density of states of these modes with $\omega<\omega_F$ is not of Debye form, $g(\omega)\propto \omega^2$. In fact, using instantaneous normal mode theory \cite{zaccone2021universal} or the $k$-gap theory of collective excitations in liquids \cite{Trachenko_2023,yu2023unveiling,brazhkin2024density}, the DOS of these overdamped modes is derived to be be linear in frequency rather than quadratic, \textit{i.e.}, $g(\omega) \propto \omega$. This linear behavior has been experimentally confirmed \cite{stamper2022experimental,jin2024temperature} (see also \cite{Ding2025}), invalidating further the assumptions made in \cite{Bolmatov2012} for shear modes with $\omega<\omega_F$. In summary, the assumptions used to derive $E_s(\omega<\omega_F)$ in the phonon theory of liquid thermodynamics \cite{Bolmatov2012} must be revisited and corrected. 

Below, we will revisit the original phonon model and propose different extensions to that. All models, will start from Eq. \eqref{toto} and differ in the description of the last term therein, related to low-frequency shear modes.
\subsection*{The original phonon theory of liquid thermodynamics} 
For clarity, we quickly summarize the fundamental steps in constructing the phonon theory of liquid thermodynamics. Starting from Eq. \eqref{toto}, we can write the phonon free energy $F_{ph}$, \cite{landau2013statistical} 
\begin{equation}
    F_{ph}=E_0+k_B T \sum_i \log \left(1-e^{-\hbar \omega_i/k_B T}\right)
\end{equation}
with $E_0$ the energy of zero-point vibrations. We then model the phonon energy, by considering the anharmonic effects related to thermal expansion, 
\begin{equation}
    E_{ph}=F_{ph}-T\frac{dF_{ph}}{dT}.
\end{equation}
By then employing a quasi-harmonic approximation, $d\omega/dT=- \alpha \omega/2$ (with $\alpha$ the thermal expansion coefficient), and by converting the discrete sum into a continuous integral over the density of states $g(\omega)$, we obtain
\begin{equation}
E_{ph}=E_0+\left(1+\frac{\alpha T}{2}\right)\int \frac{\hbar \omega}{e^{\hbar \omega/k_B T}-1}\,g(\omega) \,d\omega.
\end{equation}
At this point, one assumes a Debye density of states for both longitudinal and shear modes,
\begin{align}
g_l(\omega)=\frac{3N}{\omega_D^3}\,\omega^2,\qquad g_s(\omega)=\frac{6N}{\omega_D^3}\,\omega^2,\label{debye}
\end{align}
where the different normalization accounts for $N$ longitudinal modes and $2N$ shear modes. Moreover, $\omega_D$ is the Debye frequency that serves as an upper cutoff for the frequency. At this point, the first term in Eq. \eqref{toto} is computed using $g_l(\omega)$ and integrating between $0$ and $\omega_D$, the second term by using $g_s(\omega)$ and integrating between $\omega_F$ and $\omega_D$, and finally the last term by using $g_s(\omega)$ and integrating between $0$ and $\omega_F$. Finally, be neglecting the zero-point energy because of their weak temperature dependence that would become important only at very low temperatures, one obtains the final result:
\begin{equation}
    E= N T \left(1 +\frac{\alpha T}{2}\right)\,\left[3 \mathcal{D}\left(\frac{\hbar \omega_D}{T}\right)-\left(\frac{\omega_F}{\omega_D}\right)^3 \mathcal{D}\left(\frac{\hbar \omega_F}{T}\right)\right],\label{TB}
\end{equation}
with $\mathcal{D}(x)$ the Debye function defined as
\begin{equation}
    \mathcal{D}(x)\equiv \frac{3}{x^3}\int_0^x \frac{z^3 dz}{\exp(z)-1}.
\end{equation}
At this point, the heat capacity can be directly derived using the thermodynamic identity
\begin{equation}
    c_v=\frac{1}{N}\frac{dE}{dT}.
    \end{equation}
The harmonic crystal result can be derived by setting $\alpha=0$ and $\omega_F=0$ in Eq. \eqref{TB}. In the rest of the manuscript, we will refer to expression \eqref{TB} as the `\textit{phonon model}'.
\subsection*{Gas-like kinetic excitations}
A first extension of the original theory, proposed by Chen \cite{chen2022perspectives}, is to consider the low-frequency shear modes with $\omega<\omega_F$ as gas-like degrees of freedom contributing to the liquid energy only with their kinetic term. This is equivalent to assuming that the last term in Eq. \eqref{toto} coincides with
\begin{equation}
E_s(\omega<\omega_F)= N(\omega<\omega_F) \frac{k_B T}{2},\label{chen}
\end{equation}
where $N(\omega<\omega_F)$ is simple the number of those modes. This  is obtained by computing the complementary number of high-frequency modes $N(\omega>\omega_F)=\int_{\omega_F}^{\omega_D} g_s(\omega) d\omega$ and then by using the normalization condition $N(\omega>\omega_F)+N(\omega<\omega_F)=2N$. In the rest of the manuscript, we will refer to this approach, Eq. \eqref{chen}, as the `\textit{gas-like model}'.

We anticipate one immediate problem with this model. In fact, using this assumption, the energy of the system when $\omega_F\rightarrow \omega_D$ will be that of an ideal gas, making the high-temperature heat capacity in this model $c_v=3/2 k_B T$. This is different from the original model in which $\omega_F \rightarrow \omega_D$ coincides with the Frenkel line and with a value of $c_v=2 k_B T$, due to the total disappearance of collective shear waves.

For completeness, we also consider an alternative model in which the contribution of shear modes with $\omega<\omega_F$ is completely ignored, \textit{i.e.},
\begin{equation}
    E_s(\omega<\omega_F)=0 \label{zero}.
\end{equation}
We will label this model as the `\textit{zero model}'.
\subsection*{Overdamped liquid-like modes}
In this section, we will propose a new model that treats the low-frequency shear modes in liquids as overdamped liquid-like excitations, and not as propagating waves. 

This approach is motivated by the dispersion relation of collective shear modes in liquids that is given by:
\begin{equation}
    \omega=-\frac{i}{2 \tau} + \sqrt{v^2 k^2-\frac{1}{4 \tau^2}},
\end{equation}
where we can recognize the momentum gap, or $k$-gap, $k_g\equiv 1/(2 v \tau)$. From this expression, it is clear that collective shear excitations with frequency below $\omega_F=1/\tau$ are not underdamped propagating waves, since $\mathrm{Im}(\omega)$ is of the same order of $\mathrm{Re}(\omega)$ or even larger. This is equivalent to stating that the DOS of these modes cannot be of Debye form at low frequency. In fact, upon some simplifying assumptions, one can derive \cite{Trachenko_2023,yu2023unveiling,brazhkin2024density} that
\begin{equation}
    g_s(\omega)=\mathcal{A}\, \omega^2\,\sqrt{1+\frac{1}{4 \tau^2\omega^2}}, \label{dos}
\end{equation}
with $\mathcal{A}$ a normalization factor that is determined by the normalization condition $\int_0^{\omega_D} g_s(\omega) d\omega=2N$.

For $\omega\gg \omega_F$, Eq. \eqref{dos} reproduces the well-known Debye result, Eq. \eqref{debye}. On the other hand, for $\omega \ll \omega_F$, one obtains a linear in frequency DOS
\begin{equation}
    g_s(\omega) \approx \frac{\mathcal{A}}{2 \tau} \omega+ \dots
\end{equation}
in agreement with recent experimental results \cite{stamper2022experimental}.

We then propose to use the complete liquid density of states, Eq. \eqref{dos}, for the collective shear modes, and compute the second and third terms in the energy (Eq. \eqref{toto}) using that expression rather than the Debye DOS, that is invalid for $\omega<\omega_F$. On the other hand, the treatment of the longitudinal modes remains the same as in the original phonon model. For reference, we will label this approach as the `\textit{liquid-like model}'. We emphasize that this model, differently from the gas-like model and zero model, displays the same low-temperature and high-temperature limits as in the original formulation. Also, in order to make concrete predictions, the liquid-like approach requires the same amount of information as in the original phonon model, hence, it does not undermine its simplicity.
\subsection*{Intramolecular vibrations}
For a monatomic fluid, its kinetic energy is solely given by translational kinetic energy $c_v=c_v^{\text{trans}}$. More in general, the kinetic energy of molecular fluids consists of three contributions: the translational kinetic energy, the rotational kinetic energy and the vibrational kinetic energy. However, the original phonon model of liquid thermodynamics considers only the translational kinetic energy, disregarding the rotational and vibrational kinetic energy components. 

When operating above room temperature, the rotational degrees of freedom becomes fully excited, reaching the classical regime. In the case of linear molecules, the contribution of rotational motion to the heat capacity, $c_v^{\text{rot}}$, is equal to $R$. As for nonlinear molecules, $c_v^{\text{rot}}$, is given by $3/2R$.

The contribution of vibrational kinetic energy to heat capacity is still of quantum nature even at high temperatures due to its significant energy level spacing. Hence, within a broad temperature range, the contribution of vibrations to the heat capacity, $c_v^{\text{vib}}$, cannot be simply derived by the equipartition theorem that is only valid in the classical limit. The latter approximation becomes valid only at extremely high temperatures. Consequently, the contribution of all vibrational degrees of freedom to the heat capacity is determined by
\begin{equation}
    c_v^{\text{vib}}=R \sum_i^n\, \left(\frac{\Theta_{v,i}}{T}\right)^2\,\frac{e^{\Theta_{v,i}/T}}{\left(e^{\Theta_{v,i}/T}-1\right)^2}\label{vibra}
\end{equation}
where $\Theta_{v,i}=\hbar \nu_i/k_B$ is the vibrational temperature of the $i$-th mode and $n$ the number of vibrational degrees of freedom.

Given the contribution of translational kinetic energy, rotational kinetic energy and vibrational kinetic energy, the heat capacity of liquids can be determined by summing these components. Eq. \eqref{vibra} shows that the contribution of intramolecular motion to the heat capacity increases with temperature, while the contribution of translational motion to the liquid's heat capacity decreases with temperature.  
\begin{figure*}[htb]
    \centering
    \includegraphics[width=0.47\linewidth]{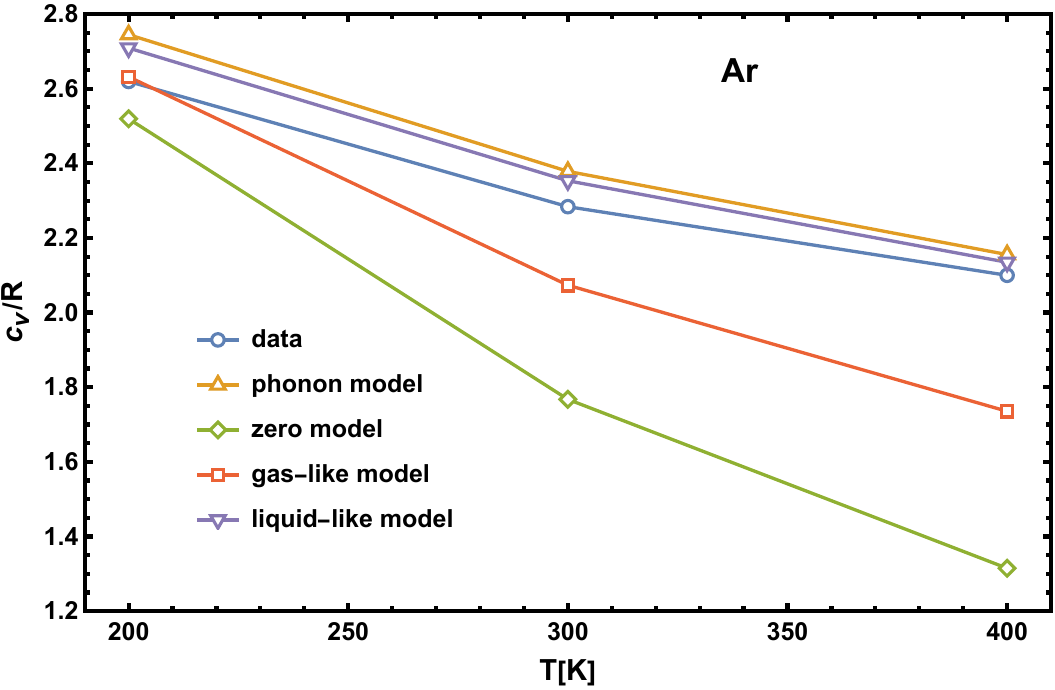}\quad 
    \includegraphics[width=0.47\linewidth]{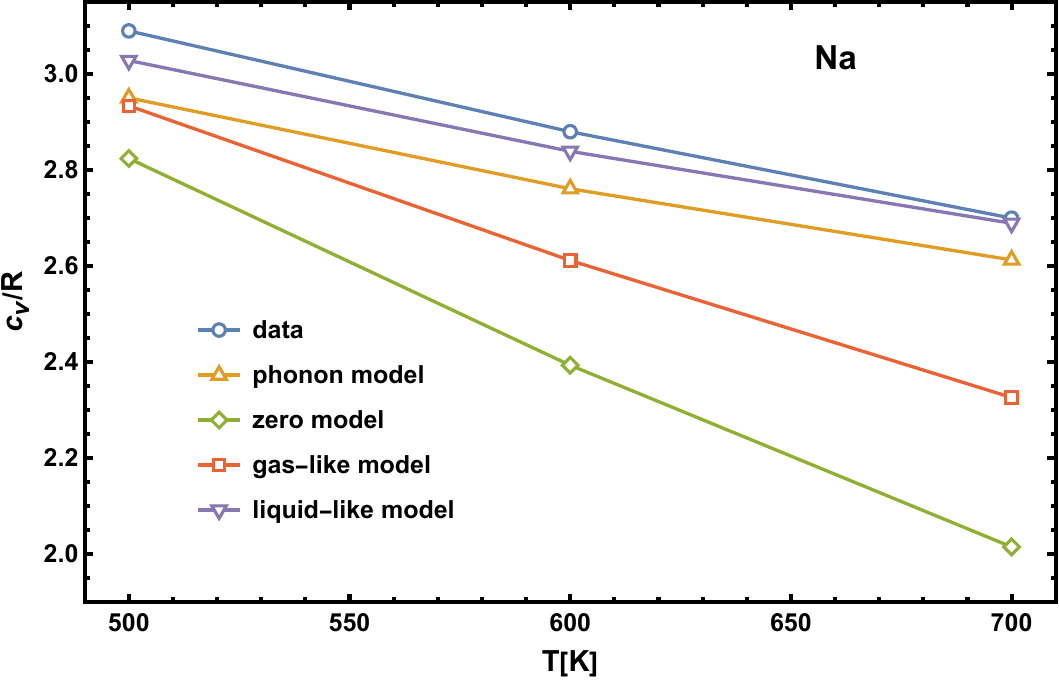}

    \vspace{0.3cm}
    
    \includegraphics[width=0.47\linewidth]{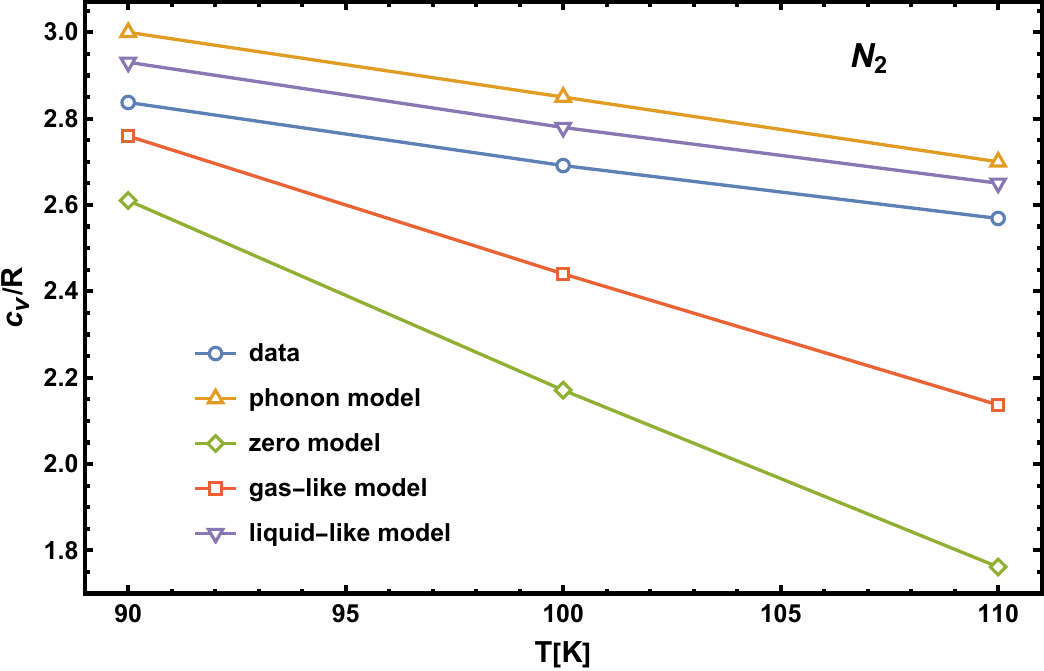}\quad
    \includegraphics[width=0.47\linewidth]{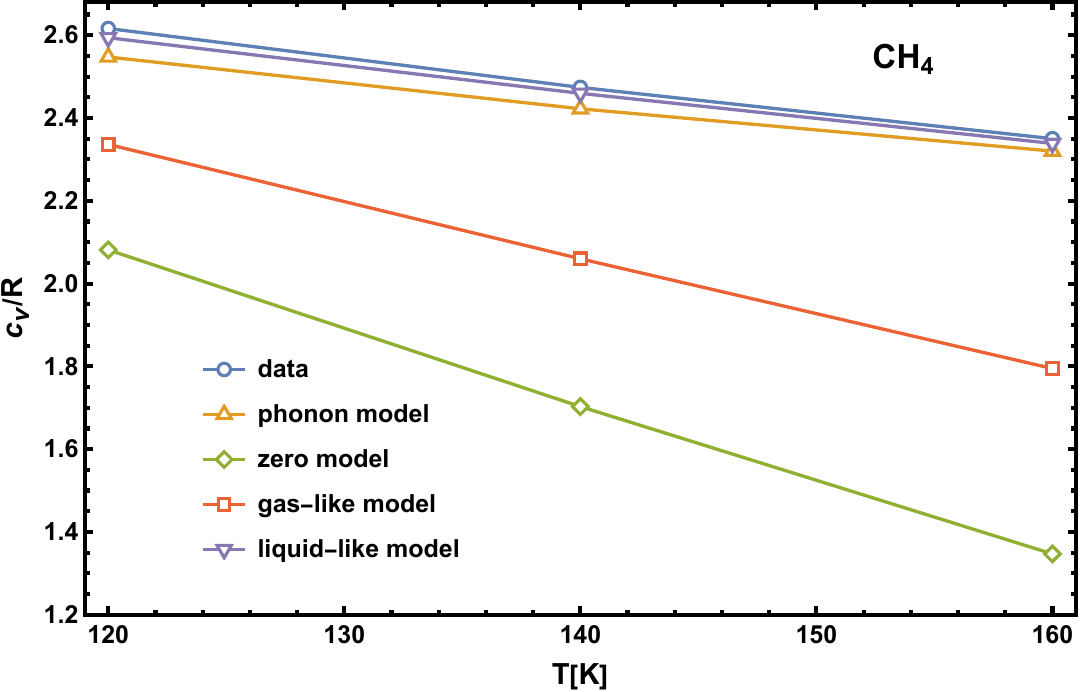}
    \caption{Experimental and calculated $c_v$ for four selected liquids Ar, Na, N$_2$ and CH$_4$. Note that for the molecular liquids (N$_2$ and CH$_4$), $c_v$ here denotes only the intermolecular part of the total heat capacity. Experimental data are taken from the NIST REFPROP database \cite{nist}. The theoretical predictions include the four different models described in the previous section.}
    \label{fig2}
\end{figure*}

As we will see, considering vibrational contributions to the liquid energy is fundamental to rationalize the behavior of liquids like CO$_2$ or C$_5$H$_{12}$ (see Fig. \ref{fig1}) that do not present a heat capacity monotonically decaying with temperature.
\section*{Comparison to experimental data}
Four models for the liquid heat capacity have been outlined in the previous section: (I) the original phonon model, (II) the gas-like model, (III) the zero-model and (IV) the liquid-like model. We now aim to compare their predictions with experimental data. For doing so, we have collected experimental data for the heat capacity of $21$ different liquids in a large range of temperatures between $30$ K and $1000$ K (see Fig. \ref{fig1}). The data are taken from the NIST REFPROP database \cite{nist}.

Before moving to the results of this comparison, let us outline in detail how the theoretical predictions have been performed. The theoretical models are based on the following physical parameters:
\begin{enumerate}
    \item The Debye frequency $\omega_D$.
    \item The Frenkel frequency $\omega_F$ or equivalently the relaxation time $\tau$. This is assumed to coincide with the Maxwell relaxation time $\tau_M\equiv \eta/G_\infty$.
    \item The thermal expansion coefficient $\alpha$.
    \item The vibrational frequencies $\nu_i$ (for molecular liquids).
\end{enumerate}
The Debye frequency is obtained from the Debye temperature of each system, $\Theta_D=\hbar \omega_D/k_B T$. In order to compute the Maxwell time, the viscosity is taken from the NIST database \cite{nist} and the infinite frequency shear modulus $G_\infty$ is calculated using the framework outlined in \cite{Keshavarzi2004}. Finally, the thermal expansion coefficient $\alpha$ is taken from \cite{Bolmatov2012} and the vibrational frequencies $\nu_i$ from \cite{wark1999thermodynamics}.

In Fig. \ref{fig2}, we report the results of our comparison for four different liquids (Ar, Na, N$_2$, CH$_4$), including two molecular liquids (N$_2$, CH$_4$). 

We first notice that the predictions from the different models present stronger deviations in the high-temperature regime. This is simply rooted in the difference between these approaches that is the nature of the shear modes below the Frenkel frequency. The number of those modes clearly increases with temperature when the gap of collective shear waves becomes larger. As a direct consequence, at low temperature these modes are few and their contribution to the heat capacity is minor. Hence, it is not surprising that the different theoretical predictions result closer to each other at low temperatures. The deviations become important by increasing temperature since the low-frequency shear modes contribute more and more to the liquid heat capacity. 

From the comparison in Fig. \ref{fig2}, it emerges that, among the various theoretical options, the zero model and the gas-like model are the ones in worst agreement with the experimental data. This is specially evident at high-temperatures where the error made by these predictions is very large. We also notice that the predictions from these models can incur in values of $c_v$ at high temperatures that are smaller than $2 R$. This is problematic since the heat capacity of a liquid is not expected to be smaller than such a value, that is indeed often taken as a criterion for the liquid-like to gas-like crossover in the supercritical region \cite{Trachenko_2016}. This is a direct consequence of treating the shear modes with $\omega<\omega_F$ as gas-like degrees of freedom, or even worse ignoring them completely. The dynamics in gases and liquids are profoundly different; hence, the prediction of the liquid heat capacity using an idealized gas-like framework incurs in several problems and inaccurate predictions.

\begin{figure}
    \centering
    \includegraphics[width=\linewidth]{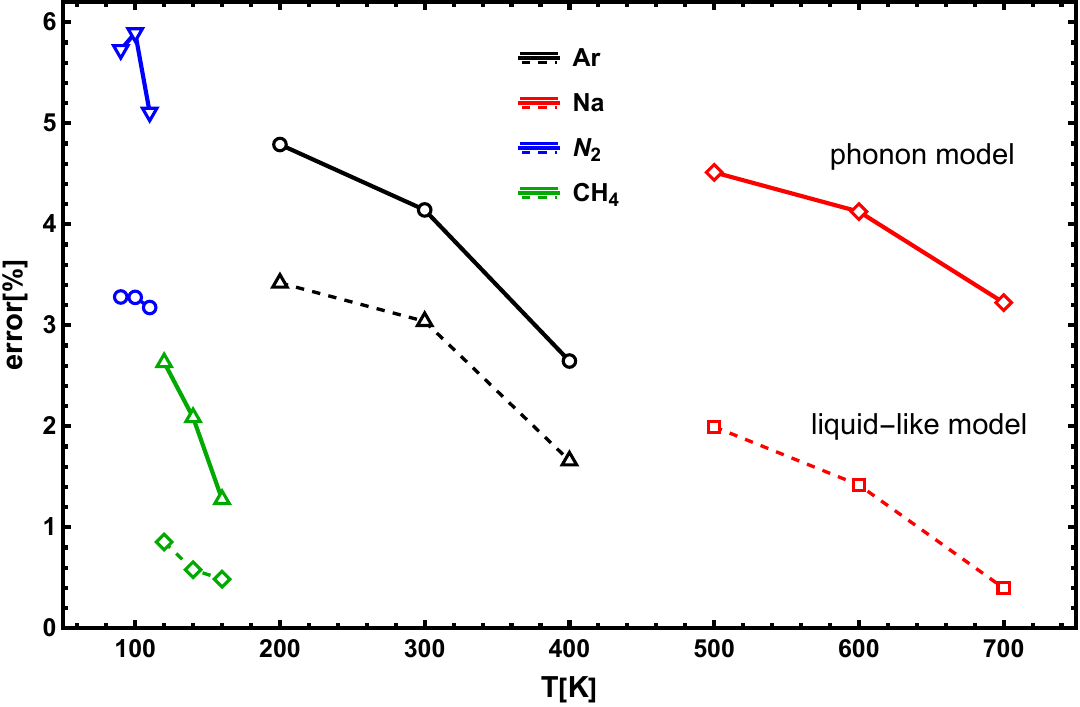}
    \caption{Percentual error of the theoretical estimates based on the original phonon model (solid lines) and the improved liquid-like model (dashed lines). Different colors correspond to different liquid. The original data are presented in Fig. \ref{fig2}.}
    \label{fig3}
\end{figure}

After excluding the zero model and the gas-like model, we are now in the position to compare the accuracy of the original phonon model and the newly proposed liquid-like model. From Fig. \ref{fig2}, it is evident that these two models provides predictions close to each other that also result to be the most accurate when compared to the experimental data. In order to compare these two models more quantitatively, we define the percentual error:
\begin{equation}
    \text{error[\%]}\equiv \left|1-\frac{c_v^{\text{model}}}{c_v^{\text{exp}}}\right|.
\end{equation}
In Fig. \ref{fig3}, we show the error function as a function of temperature from the phonon model and the liquid-like model for the four liquids considered in Fig. \ref{fig2} as well. it is evident that the liquid-like model performs much better than the original phonon model, decreasing the percentual error of a factor $\approx 2$-$3$. This is not only a practical advantage, since it gives better predictions of $c_v$, but it confirms as well a fundamental point about liquid dynamics, that is the nature of the low-frequency shear modes with $\omega<\omega_F$. Those modes are not propagating waves with Debye density of states and cannot be considered in that way. They are liquid-like modes in between gas-like kinetic excitations and solid-like vibrations. Our results support the idea that they should be treated as such.

Before jumping to our conclusions, along the lines of Ref. \cite{stamper2022experimental}, we present another comparison between previous approaches and our liquid-like model for the case of liquid Ga. In particular, we focus on the value of $c_v$ at $T=340$ K and we compare the reference value from \cite{Prokhorenko2000}, the value obtained using the experimentally measured density of states \cite{stamper2022experimental} (see details below), and the predictions from the original phonon model \cite{Bolmatov2012}, the model based on instantaneous normal modes \cite{PhysRevE.104.014103}, and finally our new liquid-like model. The value using the experimentally measured density of states using the following formula:
\begin{equation}
    c_v=k_B \int_0^\infty \left(\frac{\hbar \omega}{2 k_B T}\right)^2 \sinh\left(\frac{\hbar \omega}{2 k_B T}\right)^{-2}\,g_{\text{exp}}(\omega)d\omega,\label{ddos}
\end{equation}
with $g_{\text{exp}}(\omega)$ the density of states measured via inelastic neutron scattering experiments.

\begin{table}
    \centering
    \begin{NiceTabular}{cc}
    \hline
     Ga ($340$ K)& $c_v$ (J/Kg/K)\\
     \hline
  Reference \cite{Prokhorenko2000}  & \Block[tikz={top color=blue!15}]{*-1}{}
      341  \\
  Exp. DOS \cite{stamper2022experimental} & 340  \\
  INMs \cite{PhysRevE.104.014103} & 352 \\
  Phonon model \cite{Bolmatov2012} & 359\\
  Liquid-like model (\color{blue}this work\color{black}) & 342\\
  \hline
\end{NiceTabular}
    \caption{Heat capacity $c_v$ for liquid Ga at $340$ K: comparing the reference experimental values and the predictions from the different theoretical models.}
    \label{tab}
\end{table}

As shown in Table \ref{tab}, also in this case, the newly proposed liquid-like model outperform consistently the original phonon model and other existing approaches as well, providing a very accurate estimate of $c_v$. Our results, demonstrate that treating the low-frequency shear modes as liquid-like rather than solid-like present big advantages in the prediction of $c_v$ and it appears also more correct from a theoretical and physical perspective. 

An extended analysis on $23$ typical liquids (including $4$ noble liquids, $9$ liquid metals and $10$ molecular liquids) is presented in Appendix \ref{extended} to confirm the universal validity of our theoretical model in a large set of systems.

\begin{figure}
    \centering
    \includegraphics[width=\linewidth]{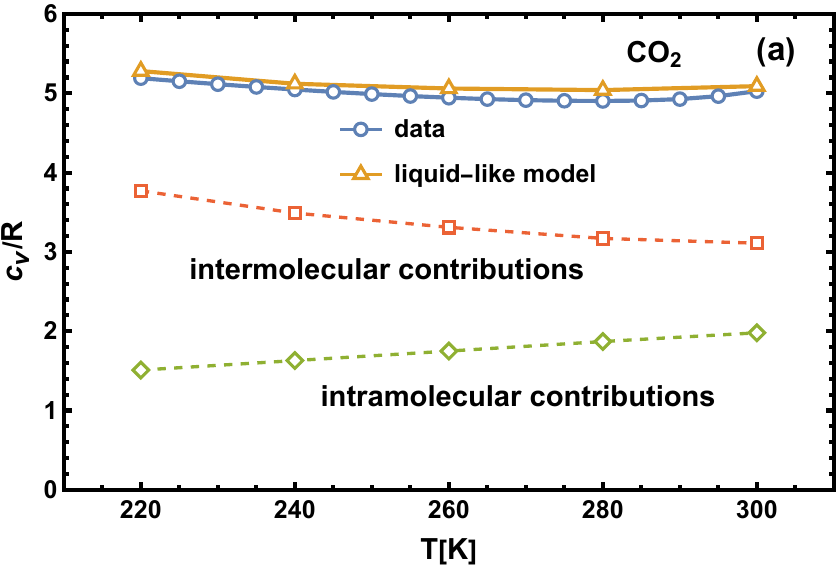}

    \vspace{0.2cm}
    
    \includegraphics[width=\linewidth]{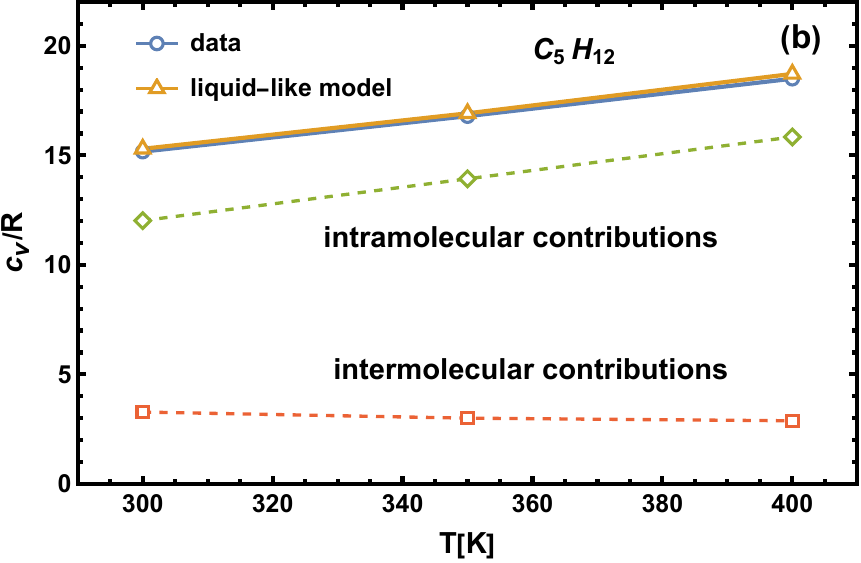}
    \caption{Comparison between the theoretical prediction of the liquid-like model with intramolecular vibrational contributions, Eq. \eqref{vibra}, and the experimental data for CO$_2$ \textbf{(a)} and C$_5$H$_{12}$ \textbf{(b)}. The red and green dashed lines correspond respectively to the intermolecular and intramolecular contributions to $c_v$.}
    \label{fig4}
\end{figure}

As already discussed above, the original phonon model of liquid thermodynamics \cite{Bolmatov2012} neglects also the role of intramolecular vibrations and consequently it is not able to reproduce a liquid heat capacity growing with temperature, as it happens in certain cases (see for example CO$_2$ and C$_5$H$_{12}$ in Fig. \ref{fig1}). In Fig. \ref{fig4}, we demonstrate that by adding this contribution, Eq. \eqref{vibra}, our liquid-like model provides accurate predictions also for the molecular liquids for which $c_v$ has a more complex trend with temperature. 

In the case of CO$_2$ (panel (a) in Fig. \ref{fig4}), the intermolecular and intramolecular contributions are of the same order of magnitude but present a very distinct temperature behavior. The intermolecular contributions decrease with temperature, while the intramolecular ones increase with $T$. Since the two terms are comparable, this produces an overall non-monotonic trend in the total heat capacity with a minimum around $\approx 260$ K, that is perfectly captured by the generalized model. On the other hand, in the case of C$_5$H$_{12}$ (panel (b) in Fig. \ref{fig4}), the intramolecular contributions that grow with $T$ are much larger than the intermolecular ones. This is the reason behind the $c_v$ monotonically increasing with $T$, in agreement with the experimental data. 
\section*{Outlook}
In this work, we have revisited the problem of predicting and modeling from theory the heat capacity of liquids, putting particular emphasis on the phonon model of liquid thermodynamics \cite{Bolmatov2012}. The starting point of our analysis relies on two important weaknesses of the original phonon model: (I) a physically incorrect interpretation of the nature of low-frequency shear modes (with frequency below the Frenkel value) and (II) the simplifying assumption of neglecting intramolecular vibrational contributions to the liquid energy.

\begin{figure}
    \centering
\includegraphics[width=\linewidth]{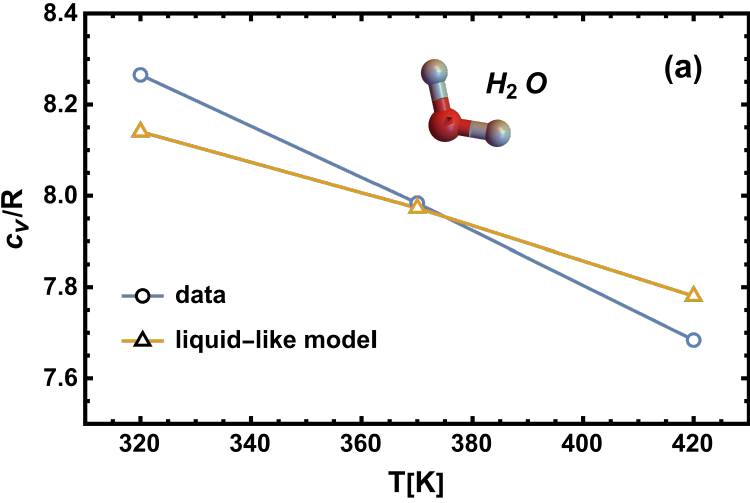}

  \vspace{0.2cm}
  \includegraphics[width=\linewidth]{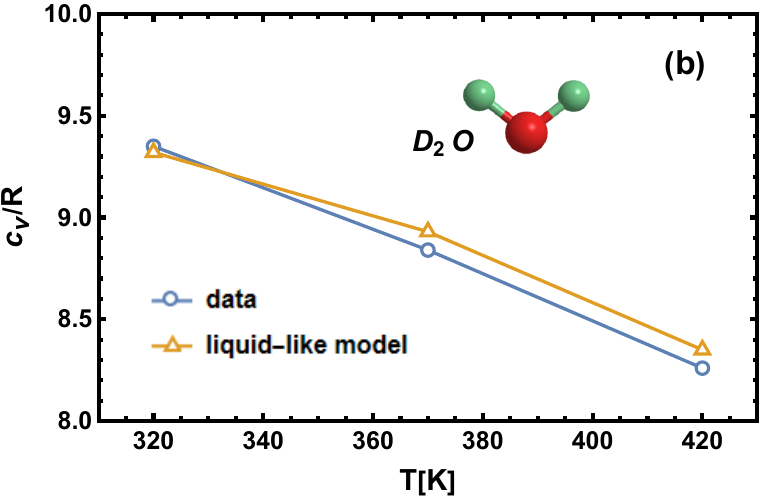}
    \caption{\textbf{(a)} Heat capacity of water: comparison between the experimental data and the theoretical prediction of the liquid-like model. \textbf{(b)} Heat capacity of heavy water: comparison between the experimental data and the theoretical prediction of the liquid-like model.}
    \label{fig5}
\end{figure}

Inspired by recent developments on the low-frequency density of states of liquids, we have proposed a new model, that is a direct extension of the original framework proposed by Bolmatov et al. \cite{Bolmatov2012}, in which the low-frequency shear modes are treated as liquid-like overdamped degrees of freedom rather than propagating solid-like waves. We have shown that this better motivated assumption leads to a stronger agreement with the experimental data, reducing the percentual error consistently.

Furthermore, we have completed this newly developed model by adding also the contributions from intramolecular vibrations that play a fundamental role in molecular liquids. By doing so, we have demonstrated that the present model is able to capture also the temperature trends of $c_v$ in complex liquids, in which the heat capacity is not a monotonically decreasing function of $T$. We notice that this behavior cannot be rationalized with the original phonon model \cite{Bolmatov2012} that predicts always a $c_v$ decreasing with $T$.

Moreover, one could ask whether this theoretical model is efficient in predicting the heat capacity of water that is probably the most complex liquid in nature. In Fig. \ref{fig5}(a), we compare the experimental data for H$_2$O between $320$ K and $420$ K with the predictions of our liquid-like model. In this case, the theoretical model appears less accurate in predicting the experimental trend. These deviations might be caused by the strong configurational contributions to the water heat capacity. Similar situations may also occur in molecules with strong hydrogen bonds, such as NH$_3$ and HF. Introducing a configurational term to the theoretical model is an interesting open question that is left for future research. For completeness, in Fig. \ref{fig5}(b), we also show the experimental data for heavy water and the predictions from our theoretical model. The two are in good agreement. The heat capacity of D$_2$O is larger than that of water and the theoretical model correctly captures this trend (\textit{isotope effect}).

Finally, it would be interesting to extend our analysis for water at higher temperature where several features, including a minimum at around $600$ K, appear. At present, this model is not able to capture those characteristics, but an extension of it might be able to. We leave this task for future research.

\section*{Acknowledgments}
We thank A. Zaccone, J. Moon, G. Chen, K. Trachenko and D. Bolmatov for useful discussions and comments on a first version of this manuscript. YL acknowledges the support of the National Natural Science Foundation of China (Grant No. 52106218), Natural Science Foundation of Chongqing (Grant No. CSTB2024NSCQ-MSX0958).
MB acknowledges the support of the Foreign Young Scholars Research Fund Project  (Grant No.22Z033100604). MB acknowledges the sponsorship from the Yangyang Development Fund.

\appendix
\section{Extended analysis}\label{extended}
In this Appendix, we present a comparison between the experimental heat capacity and our theoretical predictions for $23$ typical liquids (including $4$ noble liquids, $9$ liquid metals and $10$ molecular liquids).
\begin{table}[h]
    \centering
    \begin{NiceTabular}{cccccc}
    \hline
    \multicolumn{3}{c}{ Ar} & \multicolumn{3}{c}{Kr} \\ 
    \hline 
    T (K) & $c_v/R$ (exp) & $c_v/R$ (th) & T (K) & $c_v/R$ (exp) & $c_v/R$ (th)\\
    \hline
    200 & 2.62 & 2.71&130&2.49&2.52\\
    \hline
   300  & 2.28 &2.35&140&2.41&2.45\\
    \hline
   400  & 2.10 &2.13&150&2.35&2.38\\
     \hline
     \multicolumn{3}{c}{Ne} & \multicolumn{3}{c}{Xe} \\ 
    \hline 
     T (K) & $c_v/R$ (exp) & $c_v/R$ (th) & T (K) & $c_v/R$ (exp) & $c_v/R$ (th)\\
    \hline
 30   & 2.25
&2.23
&200&2.45
&2.50
\\
    \hline
    35 & 2.15&2.17&220&2.34&2.40\\
    \hline
   40  & 2.07&2.12&240&2.26&2.31\\
     \hline
\end{NiceTabular}
    \caption{Experimental (exp) and calculated (th) heat capacity $c_v$ for noble liquids.}
    \label{tab2}
\end{table}
\begin{table}[h]
    \centering
    \begin{NiceTabular}{cccccc}
    \hline
    \multicolumn{3}{c}{Cs} & \multicolumn{3}{c}{Ga} \\
    \hline
    T (K) & $c_v/R$ (exp) & $c_v/R$ (th) & T (K) & $c_v/R$ (exp) & $c_v/R$ (th)\\
    \hline
    400 &3.20&3.14&400&2.90&2.84\\
    \hline
    500& 3.00&2.96&500&2.78&2.76\\
    \hline
    600& 2.80&2.78&600&2.69&2.68\\
     \hline
     700& 2.63&2.61&700&2.61&2.61\\
     \hline
     \multicolumn{3}{c}{Hg} & \multicolumn{3}{c}{In} \\
    \hline
     T (K) & $c_v/R$ (exp) & $c_v/R$ (th) & T (K) & $c_v/R$ (exp) & $c_v/R$ (th)\\
    \hline
    300& 2.90&2.88&500&3.06&3.02\\
    \hline
    400& 2.72&2.71&600&2.96&2.95\\
    \hline
    500& 2.60&2.59&700&2.87&2.86\\
     \hline
     600& 2.54&2.51&800&2.80&2.79\\
     \hline
     \multicolumn{3}{c}{K} & \multicolumn{3}{c}{Na} \\
    \hline
     T (K) & $c_v/R$ (exp) & $c_v/R$ (th) & T (K) & $c_v/R$ (exp) & $c_v/R$ (th)\\
    \hline
    500& 2.99&2.99&500&3.09&3.03\\
    \hline
    600& 2.80&2.82&600&2.88&2.84\\
    \hline
    700& 2.65&2.68&700&2.70&2.69\\
     \hline
      800& 2.52&2.54&-&-&-\\
     \hline
     \multicolumn{3}{c}{Pb} & \multicolumn{3}{c}{Rb} \\
    \hline
     T (K) & $c_v/R$ (exp) & $c_v/R$ (th) & T (K) & $c_v/R$ (exp) & $c_v/R$ (th)\\
    \hline
    700& 2.90&3.01&500&3.00&2.97\\
    \hline
    800& 2.80&2.87&600&2.80&2.79\\
    \hline
    900& 2.70&2.74&700&2.64&2.64\\
     \hline
     1000& 2.60&2.63&800&2.50&2.49\\
     \hline
     \multicolumn{3}{c}{Sn} & \multicolumn{3}{c}{-} \\
    \hline
     T (K) & $c_v/R$ (exp) & $c_v/R$ (th) & T (K) & $c_v/R$ (exp) & $c_v/R$ (th)\\
    \hline
    600& 3.00&2.98&-&-&-\\
    \hline
    700& 2.90&2.89&-&-&-\\
    \hline
    800& 2.81&2.81&-&-&-\\
     \hline
     
\end{NiceTabular}
    \caption{Experimental (exp) and calculated (th) heat capacity $c_v$ for liquid metals.}
    \label{tab3}
\end{table}
\begin{table}[h]
    \centering
    \begin{NiceTabular}{cccccc}
    \hline
    \multicolumn{3}{c}{CH$_4$} & \multicolumn{3}{c}{CO} \\
    \hline
    T (K) & $c_v/R$ (exp) & $c_v/R$ (th) & T (K) & $c_v/R$ (exp) & $c_v/R$ (th)\\
    \hline
     120& 2.62&2.59&95&2.88&2.84\\
    \hline
    140& 2.47&2.46&105&2.72&2.70\\
    \hline
    160& 2.35&2.34&115&2.59&2.58\\
     \hline
     \multicolumn{3}{c}{F$_2$} & \multicolumn{3}{c}{H$_2$S} \\
    \hline
     T (K) & $c_v/R$ (exp) & $c_v/R$ (th) & T (K) & $c_v/R$ (exp) & $c_v/R$ (th)\\
    \hline
    60& 3.12&3.11&250&3.19&3.18\\
    \hline
    70& 2.94&2.91&300&2.83&2.83\\
    \hline
    80& 2.82&2.79&350&2.59&2.59\\
     \hline
     \multicolumn{3}{c}{N$_2$} & \multicolumn{3}{c}{O$_2$} \\
    \hline
     T (K) & $c_v/R$ (exp) & $c_v/R$ (th) & T (K) & $c_v/R$ (exp) & $c_v/R$ (th)\\
    \hline
    90& 2.84&2.93&100&2.72&2.65\\
    \hline
    100& 2.69&2.78&110&2.60&2.55\\
    \hline
    110& 2.57&2.65&120&2.50&2.49\\
     \hline
     \multicolumn{3}{c}{CO$_2$} & \multicolumn{3}{c}{C$_5$H$_{12}$} \\
    \hline
     T (K) & $c_v/R$ (exp) & $c_v/R$ (th) & T (K) & $c_v/R$ (exp) & $c_v/R$ (th)\\
    \hline
    220& 3.68&3.77&300&3.15&3.28\\
    \hline
    240& 3.42&3.49&350&2.87&3.00\\
    \hline
    260& 3.19&3.31&400&2.66&2.88\\
     \hline
     280& 3.04&3.17&-&-&-\\
     \hline
     300& 3.05&3.11&-&-&-\\
     \hline
     \multicolumn{3}{c}{H$_2$O} & \multicolumn{3}{c}{D$_2$O} \\
    \hline
     T (K) & $c_v/R$ (exp) & $c_v/R$ (th) & T (K) & $c_v/R$ (exp) & $c_v/R$ (th)\\
    \hline
    320& 2.58&2.52&320&3.02&3.01\\
    \hline
    370& 2.40&2.39&370&2.69&2.73\\
    \hline
    420& 2.20&2.25&420&2.31&2.35\\
     \hline
\end{NiceTabular}
    \caption{Experimental (exp) and calculated (th) heat capacity $c_v$ for molecular liquids (intermolecular part).}
    \label{tab3}
\end{table}

\end{document}